\documentclass[twocolumn,showpacs,preprintnumbers,amsmath,amssymb]{revtex4}

\usepackage{graphicx}
\usepackage{dcolumn}
\usepackage{bm}

\begin{document}
\title{Oxygen phonon branches in overdoped La$_{1.7}$Sr$_{0.3}$CuO$_4$}
\author{L. Pintschovius$^{1}$, D. Reznik$^{1,2}$,  and K. Yamada$^{3}$}
\affiliation{$^1$Forschungszentrum Karlsruhe, Institut f\"ur Festk\"orperphysik, P.O.B. 3640, D-76021 Karlsruhe, Germany\\
$^2$Laboratoire L$\acute{e}$on Brillouin, CE Saclay, F-91191
Gif-sur-Yvette, France\\ $^3$Institute for Materials Research
(IMR),
Tohoku University, Sendai 980-8577, Japan}
\email{pini@ifp.fzk.de}
%

%
\date{\today}

\begin{abstract}
The dispersion of the Cu-O bond-stretching vibrations in overdoped
La$_{1.7}$Sr$_{0.3}$CuO$_4$ (not superconducting) has been
studied by high resolution inelastic neutron scattering. It was
found that the doping-induced renormalization of the so-called
breathing  and the half-breathing modes  is larger than in
optimally doped La$_{1.85}$Sr$_{0.15}$CuO$_4$. On the other hand,
the phonon linewidths are generally smaller in the overdoped
sample.  Features observed in optimally doped
La$_{1.85}$Sr$_{0.15}$CuO$_4$ which suggest a tendency towards
charge stripe formation are absent in overdoped
La$_{1.7}$Sr$_{0.3}$CuO$_4$.
\end{abstract}

\maketitle

\section{Introduction}
It has been known for quite some time that the frequencies of the
Cu-O bond-stretching vibrations are strongly renormalized upon
doping in all the cuprates investigated so far \cite{review}. The
frequency  renormalization and a concurrent increase of the
phonon linewidths are clear evidence of a strong coupling of
these phonon modes to the charge carriers. Renewed interest in
this phenomenon was generated by the observation of a kink in the
quasi-particle dispersion by angle-resolved photoemission (ARPES)
data \cite{Lanzara} at an energy corresponding to the
bond-stretching phonon frequencies. Though the interpretation of
the ARPES data remains a matter of debate, it is now widely
acknowledged that phonons might play an important role for the
electron dynamics in the high temperature superconductors. For a
brief moment, there seemed to be even a close correlation between
the frequency renormalization and the superconducting transition
temperature \cite{Fukuda0}. The evidence came from inelastic
x-ray scattering (IXS) data on overdoped
La$_{1.71}$Sr$_{0.29}$CuO$_4$. However, later measurements of the
same kind \cite{Fukuda1} did not confirm the early data and
therefore, the original paper \cite{Fukuda0} has been retracted.
The recent IXS measurements revealed that the phonon softening in
the (100)-direction starts to saturate around optimal doping (x =
0.15). In this paper, we present high-quality inelastic neutron
scattering data on overdoped, non-superconducting
La$_{1.7}$Sr$_{0.3}$CuO$_4$ for both the (100)- and the
(110)-directions allowing for a detailed comparison of both
frequencies and line-widths with corresponding data for optimally
doped samples. We show that the phonon softening on going from
optimally doped  to  overdoped LSCO is weak in the (100)
direction but quite strong in the (110) direction. Further, we
demonstrate that all signatures reflecting dynamic charge
inhomogeneities seen in compounds with doping levels about 1/8 as
well as  in optimally doped LSCO \cite{nature} are absent in
La$_{1.7}$Sr$_{0.3}$CuO$_4$.

\section{Experimental}

The sample consisted of a single crystal of composition
La$_{1.7}$Sr$_{0.3}$CuO$_4$ and a volume of about 1 cm$^3$. Its
mosaic spread was about 1 degree.  The experiments were carried
out on the triple-axis spectrometer 1T located at the
Orph$\acute{e}$e reactor using doubly focusing monochromator
(Cu220) and analyzer (PG002) crystals. Cu220 was used as
monochromator to achieve high resolution. The actual resolution -
depending on the focusing conditions - is indicated in the lower
panels of Fig. 2. Measurements were carried out both in the
100-001 and in the 100-010  scattering planes. All measurements
were performed at T=10 K.

\begin{figure}
\centerline{\includegraphics[height=3.0in]{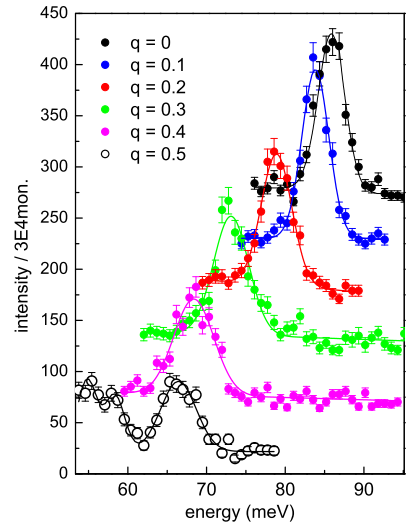}}
\caption{(Color online) Energy scans taken along the line Q =
(5-q,0,0) at T = 10 K. Successive scans were off-set by 50 counts
for the sake of clarity. Lines depict fit curves. The peaks are
associated with longitudinal Cu-O bond-stretching vibrations,
except for the double peak at {\bf q} = 0.5 below 60 meV which is
related to Cu-O bond-bending vibrations.} \label{firstfigure}
\end{figure}

\section{Experimental results}
\subsection{Plane-polarized Cu-O bond-stretching modes}
\begin{figure}
\centerline{\includegraphics[height=4in]{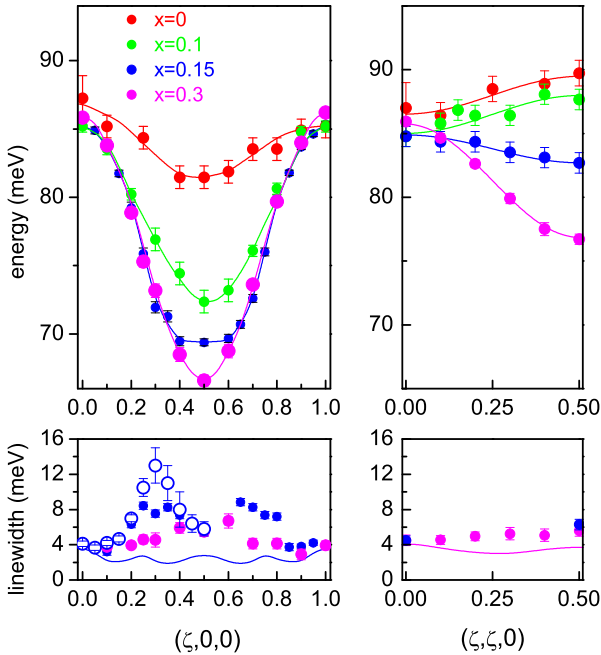}}
\caption{(Color online) Upper panels: Dispersion of the
longitudinal high-energy phonon branches in the (100)- and in the
(110)-directions, respectively, for various doping levels in
La$_{2-x}$Sr$_{x}$CuO$_4$. Lines are a guide to the eye. Data for
x = 0 and 0.1 were taken from \cite{review}. Data for x = 0.15
were taken from  \cite{pinibraden}. Lower panels: The data points
show the line-widths (full width at half maximum) of the phonon
peaks and the lines show the experimental resolution. The
wavyness of the lines is due to focusing effects. Recent data
\cite{JLTP}  for x=0.15 measured with an improved q-resolution in
the transverse direction are included as open symbols. The
displacement patterns of the zone boundary modes are shown in
Fig. 4.} \label{secondfigure}
\end{figure}

Fig. 1 demonstrates the high quality of the raw data. The
dispersion and the line-widths of the high-energy longitudinal
modes in the (100) and in the (110) directions are depicted in
Fig. 2.  Obviously, the strong doping-induced frequency changes
between undoped and optimally doped LSCO are further enhanced
when going to overdoped samples, in particular in the (110)
direction. We emphasize that the doping-induced softening is
observed only in the longitudinal branches: the transverse branch
in the (100) direction was found to be completely flat as for all
other LSCO samples investigated previously, and the dispersion of
the transverse modes in the (110) direction is even slightly
reduced with increasing doping levels (Fig. 3). As was stated
previously \cite{chaplot}, the pronounced dispersion of the
transverse modes in the (110)-direction can be completely
accounted for by Coulomb forces and is not indicative of a strong
electron-phonon coupling. This view is further corroborated by
the narrow line-widths of the transverse modes which are much
smaller than those of the longitudinal ones. Examples are shown
for zone boundary modes in Fig. 5.

\begin{figure}
\centerline{\includegraphics[height=2.5in]{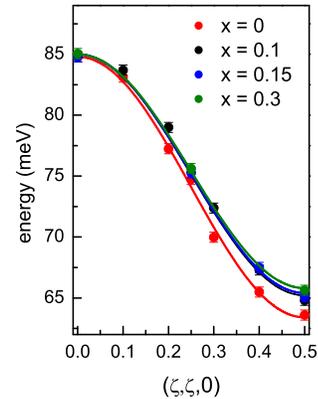}}
\caption{(Color online) Dispersion of the transverse high-energy
phonons in the (110)-direction for various doping levels in
La$_{2-x}$Sr$_{x}$CuO$_4$. Lines are a guide to the eye. Data for
x $<$ 0.3 were taken from \cite{review}. The displacement pattern
of the zone boundary mode is shown in Fig. 4 (middle).}
\label{thirdfigure}
\end{figure}

\begin{figure}
\centerline{\includegraphics[height=2.5in]{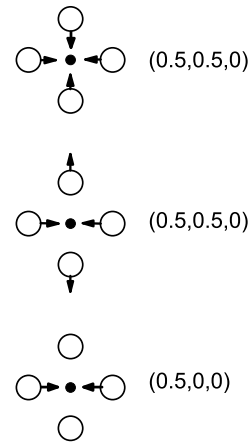}}
\caption{Displacement patterns of zone-boundary bond-stretching
modes. Top: longitudinal mode in the (110)-direction (breathing
mode); middle: transverse mode in the (110)-direction quadrupolar
mode); bottom: longitudinal mode in the (100)-direction
(half-breathing mode). Circles and full points denote oxygen
atoms and copper atoms, respectively. Only the displacements in
the Cu-O planes are shown. All other displacements are small for
these modes.} \label{fourthfigure}
\end{figure}

\begin{figure}
\centerline{\includegraphics[height=2.5in]{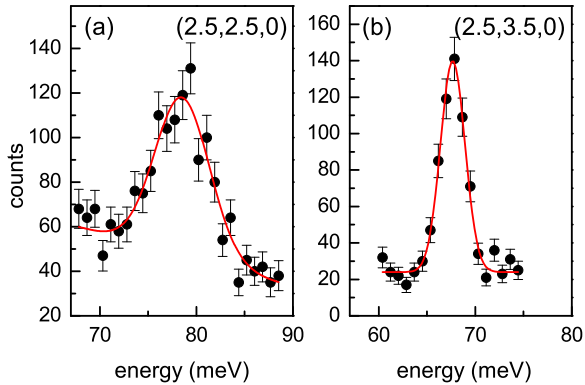}}
\caption{(Color online) Energy scans taken at the zone boundary
in the (110)-direction at T = 10 K. The peak in (a) corresponds to
the longitudinal and in (b) to the transverse Cu-O bond-stretching
mode. The  two modes  are of breathing or quadrupolar character,
respectively. The width of the quadrupolar mode is resolution
limited. } \label{fifthfigure}
\end{figure}

\subsection{C-polarized modes}
Since previous investigations \cite{review} had shown that a
particular c-polarized oxygen mode exhibits a very strong
doping-induced softening  and acquires simultaneously  a massive
broadening, we investigated this mode as well on our overdoped
sample. This mode has been termed O$_z^z$-mode by Falter and
co-workers \cite{falter2} because it is a z(=c)-polarized zone
boundary (Z-point) mode. These authors predicted a strong
electron-phonon coupling of this mode prior to experiment based
on the following argument: the displacement pattern is such that
all apical oxygen atoms move simultaneously towards the Cu-O
planes thereby inducing strong charge fluctuations. In order to
see whether the anomalous character of this mode is enhanced or
rather reduced on overdoping we performed an energy scan at
Q=(0,0,15).  This Q-point gives the maximum inelastic structure
factor for the O$_z^z$-mode but unfortunately, another
c-polarized apical oxygen mode has a non-negligible structure
factor as well. The energy of this mode is known from
measurements in other Brillouin zones to be 56 meV. Hence, the
intensity distribution was fitted with two Gaussians
corresponding to the O$_z^z$-mode and the second mode,
respectively (Fig. 6). We found that the energy of the
O$_z^z$-mode in overdoped LSCO  is practically the same as in
optimally doped LSCO, i.e. very low compared to that in the
undoped parent compound (48 meV vs. 70 meV). The line-width
observed in overdoped LSCO is smaller than in optimally doped
LSCO but only slightly so (by about 10 $\%$).

\begin{figure}[htb]
\includegraphics[width=.3\textwidth]{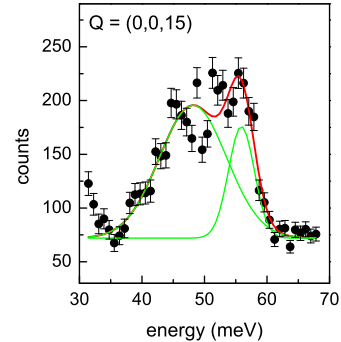}
\caption{(Color online) Energy scan taken at Q = (0,0,15) at T=10
K. The intensity distribution was fitted with two Gaussians
corresponding to the contributions of the O$_z^z$-mode (around 48
meV) and another mode of A$_g$ type (around 56 meV). The energy
resolution was about 4 meV.} \label{sixthfigure}
\end{figure}

\section{Discussion}

A comparison of the neutron results for x=0.30 with the recently
published x-ray results \cite{Fukuda1} for the longitudinal
bond-stretching modes in the (100)-direction in
La$_{1.85}$Sr$_{0.15}$CuO$_4$  shows in general very good
agreement. We note that the results for the q-dependent
line-widths depicted in Fig. 3b of the x-ray paper suggest a weak
maximum around q = 0.3 which is absent in  the neutron data. This
seems to be simply a consequence of the fact that the neutron
data show a monotonic increase of the line-width towards the zone
boundary whereas the  x-ray signal was lost in this region of
q-space. In a later inelastic x-ray study \cite{ikeuchi}, data
were published not only for the (100), but also for some
longitudinal modes in the (110) direction. Again, good agreement
is found between the x-ray data and the neutron data .

The doping-induced frequency renormalization of bond-stretching
modes in high-T$_c$ materials has been known for many years. It
bears clear resemblance to phonon anomalies found in many
conventional superconductors like, e.g., Nb$_3$Sn
\cite{pininb3sn}. Therefore, it is generally seen as evidence of
a substantial coupling of the bond-stretching modes to the
quasi-particles. Consequently, this issue has been addressed in
several theoretical papers. Falter and co-workers start from the
electronic band structure calculated within the local density
approximation (LDA)
\cite{falter2,falter3,falter4,falter5,falter6}. The failure of LDA
theory to reproduce the insulating ground-state of the undoped
parent compounds is remedied by suitably imposing the
long-wavelength limit of the electronic polarizability
\cite{falter6}. In this theory, the screening processes producing
the softening are described in terms of charge fluctuations on the
outer shells of the ions. The doping dependence of the
bond-stretching modes is well accounted for in the range from
undoped to optimally doped LSCO. In particular, the theory
correctly reproduces the much stronger softening in the (100)-
direction as compared to that in the (110)-direction for
underdoped and optimally doped samples. Unfortunately, no
predictions were made for overdoped samples.

\begin{figure}
\includegraphics[width=.3\textwidth]{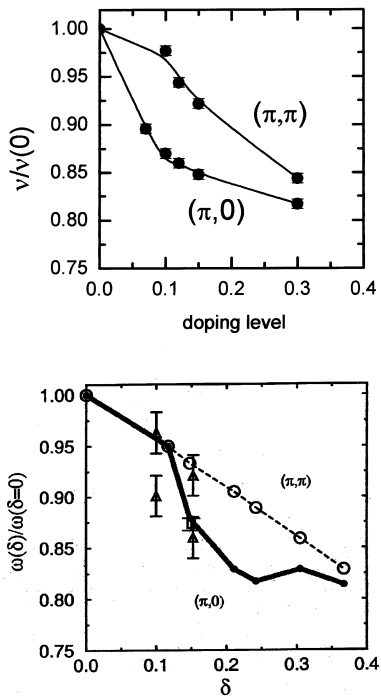}
\caption{Top: Experimentally observed frequency renormalization in
LSCO of the half-breathing mode ($\pi$,0) and of the breathing
mode ($\pi$,$\pi$) , respectively \cite{footnote1}. Data for x $<$
0.30 were taken from \cite{review,pinibraden,nature}; bottom:
Predicted doping dependence of the frequencies of the
half-breathing mode and of the breathing mode, respectively,
after Horsch and Khaliullin \cite{horsch1}. The triangles denote
experimental data known at the time of the report.}
\label{seventhfigure}
\end{figure}

Other groups have chosen a very different approach to explain the
phonon renormalization of the bond-stretching modes upon doping,
i.e. using the t-J model and extending it to explicitly include
electron-phonon couplings \cite{Roesch,horsch}. The correct
anisotropy between the [1,1] and the [1,0] directions in
optimally doped LSCO has been reported for the first time by
Khaliullin and Horsch \cite{horsch}. As is explained in
\cite{horsch,horsch2} , the rapid drop of the frequency of the
half-breathing mode when going from the undoped to the optimally
doped compound results from a polaron peak in the electron
density fluctuation spectrum N({\bf q},$\omega$) being in the
same energy range as the bond-stretching phonons. Similar results
have been reported recently by R\"osch and Gunnarsson
\cite{Roesch}. Horsch and Khaliullin have the merit of having
used their theory to predict the doping dependence of the
frequencies of the half-breathing mode and of the breathing mode
up to high doping levels, and that several years prior to the
experiments \cite{horsch1}. The agreement between theory and
experiment is impressive(Fig. 7).

In this context, we would like to mention that the downward
dispersion of the longitudinal Cu-O bond-stretching branches has
been successfully predicted by density functional theory for
another member of the cuprate family, i.e. YBa$_{2}$Cu$_3$O$_7$
(O7) \cite{bohnen}. O7 is generally considered to be slightly
overdoped. In the (100)- resp. (010)-directions, the calculated
dispersion curves agree very well with the experimental ones
observed on optimally doped O6.95 \cite{PhysicaC}. In the
(110)-direction, however, the calculated downward dispersion is
somewhat stronger than observed in experiment on a highly doped
sample \cite{reichardt}. Unfortunately, the doping evolution of
the downward dispersion cannot be studied by this theory in its
present state because of the failure to describe the insulating
state of the  parent compounds.

At first glance, the monotonic decrease of the zone boundary
frequencies with increasing doping suggests a monotonic increase
of the electron-phonon coupling strength. We note, however, that
the electron-phonon coupling strength is not directly related to
the frequency renormalization but rather to the phonon linewidths.
Therefore, the reduction of the phonon linewidths on overdoping
is direct evidence that the  coupling of the bond-stretching
modes to the quasi-particles  is indeed weaker in overdoped LSCO
when compared to optimally doped LSCO. In particular, the massive
line broadening of phonons propagating along (100) for q values
around 0.3 observed for doping values 0.07 $<$ x $<$ 0.15 is
absent in overdoped LSCO (Fig. 8). As has been explained in a
recent paper \cite{nature} based on results for compounds showing
static stripe phase order , the extremely large linewidths
observed in the (100) direction at halfway to the zone boundary
are very probably related to dynamic charge stripe formation.
Another signature of dynamic charge stripe formation discussed in
\cite{nature} is a very steep slope of the phonon dispersion
around {\bf q} = (0.25,0,0) leading to a strong deviation from a
sinusoidal shape of the phonon dispersion curves (see also Fig.
25 in  \cite{physstatsol}) In optimally doped LSCO, the maximum
slope is not  as large as found in compounds showing static
stripe order but is still relatively large \cite{footnote2}. In
overdoped LSCO, however, the bond-stretching phonon frequencies
along (100) follow a cosine behavior very well. Therefore, it can
be said that the signatures of dynamic charge stripe formation
observed in optimally and in slightly underdoped LSCO (x = 1/8)
are completely absent in a strongly overdoped (and
non-superconducting) sample. In this context, we would like to
point out that the signatures of dynamic charge stripe formation
become  weak as well when going to the strongly underdoped side
of the phase diagram. In particular, the phonon dispersion curve
along (100) becomes approximately sinusoidal (see the data for x
= 0.1 in Fig. 2 and further data for x = 0.07  in \cite{nature}).
On the other hand, the linewidths observed in a sample with
x=0.07 (T$_c$=20 K) around {\bf q} = (0.3,0,0) are still rather
high \cite{nature} which means that it is these linewidths which
are most closely correlated with the superconducting transition
temperatures \cite{footnote3}.

\begin{figure}
\centerline{\includegraphics[height=2.5in]{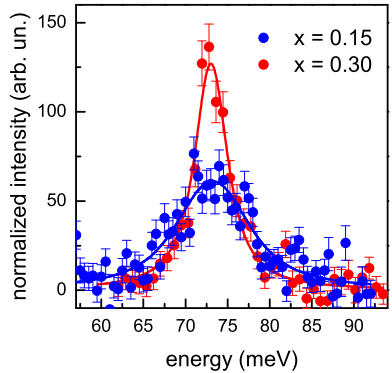}}
\caption{(Color online) Comparison of energy scans taken at Q =
(5-0.3,0,0) on optimally doped \cite{JLTP} and on overdoped LSCO
at T = 10 K. The data were corrected for background scattering and
normalized to yield the same integrated intensity. Both scans
were taken with the same experimental set-up.  Note that the
width observed on the optimally doped sample is somewhat larger
than that reported in \cite{pinibraden} because of a better
momentum resolution in the later experiment. } \label{eigthfigure}
\end{figure}

We note that none of the theories discussed above predicts any of
the signatures of dynamic stripe formation observed in
experiment. Theory did predict the tendency for charge stripe
formation in the cuprates very early \cite{Zaanen, Machida,Emery}
but failed to make detailed predictions for the doping dependence
of this phenomenon. Qualitatively speaking, stripe formation is
related to the correlated nature of the quasiparticles in the
Cu-O2 planes. Therefore, the absence of signatures of stripe
formation in heavily overdoped LSCO probably reflects the
transition from a correlated metal to a normal metal on
overdoping.

\section{Conclusions}
In summary, we have found that the doping-induced frequency
renormalization of bond-stretching modes is enhanced when going
from an optimally doped to a strongly overdoped sample. In
particular, the renormalization of the longitudinal modes in the
(110)-direction becomes quite strong on overdoping, in very good
agreement with theoretical predictions. However, the phonon
linewidths decrease at the same time indicating that the
electron-phonon coupling strength is not enhanced on overdoping.
These results mean that superconductivity is not correlated with
the general phonon renormalization but rather with the phonon
linewidths. More specifically,  the massive line broadenings
observed around {\bf q} =(0.25,0,0) in  LSCO at doping levels
close to optimum doping indicative of dynamical charge stripe
order are not found in strongly overdoped, non-superconducting
LSCO. These  linewidths are even smaller than the corresponding
ones in strongly underdoped, but still superconducting LSCO.

\section{Acknowledgements}
The work at Tohoku University was supported by grants from the
Japanese Ministry of Education, Culture, Sports, Science and
Technology.


\begin{thebibliography}{10}
\bibitem{footnote1} Note that the frequency
renormalization plotted in this figure cannot be compared
directly to the quantities plotted in the x-ray papers (fig. 4 of
\cite{Fukuda1} resp. fig. 11 of cite{ikeuchi}). The so-called
phonon softening shown in the x-ray papers is simply the
difference between the zone center and the zone boundary
frequencies, whereas the softening plotted in our fig. 7 is
calculated with respect to the undoped case. Further, our fig. 7
is based on true zone boundary frequencies and not on a fit of
the whole branch with a sine-function. It is basically this fit
with a sine-function which produces the local maximum in the
softening vs. doping curve around optimal doping in
\cite{Fukuda1} because the dispersion of the LO branch along
(100) deviates strongly from a sine-function for optimally doped
LSCO but not for  overdoped LSCO.
\bibitem{footnote2} A close inspection of Fig. 2 reveals that the
maximum slope of the dispersion curve in the (100)-direction is
about 20 \% larger in optimally doped LSCO compared to overdoped
LSCO in spite of the smaller overall dispersion. A recent
re-investigation of optimally doped LSCO with somewhat improved
momentum resolution  revealed that the maximum slope is even
larger than deduced from the first study \cite{pinibraden}, i.e.
about 50 \% larger than in overdoped LSCO, although still smaller
than in the static stripe phase
La$_{1.48}$Nd$_{0.4}$Sr$_{0.12}$CuO$_4$ \cite{JLTP}.
\bibitem{footnote3} This correlation breaks down when the crystal
structure produces a pinning potential to make the charge stripes
static as in La$_{1.48}$Nd$_{0.4}$Sr$_{0.12}$CuO$_4$: static
stripes are apparently detrimental to superconductivity
\cite{JMT}.
\end{thebibliography}
\end{document}